\def\ee{\end{equation}}
\def\be{\begin{equation}}
\def\eea{\end{eqnarray}}
\def\bea{\begin{eqnarray}}
\def\eeas{\end{eqnarray*}}
\def\beas{\begin{eqnarray*}}

\documentstyle[preprint,tighten,aps]{revtex}

\def\lapprox{\mathrel{\mathop
  {\hbox{\lower0.5ex\hbox{$\sim$}\kern-0.8em\lower-0.7ex\hbox{$<$}}}}}
\def\gapprox{\mathrel{\mathop
  {\hbox{\lower0.5ex\hbox{$\sim$}\kern-0.8em\lower-0.7ex\hbox{$>$}}}}}

\begin{document}
\author{ B. Ricci $^1$ and
        F.L. Villante $^1$}

\address{
  $^{1}$ Dipartimento di Fisica dell'Universit\`a di Ferrara and
        Istituto Nazionale di Fisica Nucleare, Sezione di Ferrara,
        via Paradiso 12, I-44100 Ferrara, Italy .
}

\preprint{\vbox{\noindent
          \null\hfill  INFNFE-05-00}}

\title{Helioseismic determination of Beryllium neutrinos produced in the Sun}

\date{May 2000}

\maketitle

\begin{abstract}
We provide a determination
of the Beryllium neutrino luminosity 
directly by means of helioseismology, without using 
additional assumptions. 
We have constructed solar models where Beryllium neutrino,
($\nu_{Be}$) production is artificially changed  by varying
in an arbitrary way the zero energy  astrophysical S-factor  $S_{34}$
for the reaction $^3{\rm He}+^4{\rm He}\rightarrow ^7{\rm Be}+ \gamma$.
 Next we have compared the properties of such models with helioseismic
determinations of photospheric helium abundance, depth of the convective
zone and sound speed profile. We find that helioseismology directly
confirms the production rate of  $\nu_{Be}$ as
predicted by SSMs to within  $\pm 25\%$ ($1\sigma$ error).
This constraint is somehow weaker than 
that  estimated from uncertainties of the SSM
($\pm 10\%$), however it relies on direct observational 
data.

\end{abstract}

\section{Introduction}
As well known, the production of neutrinos from $^7{\rm Be}+e^{-}\rightarrow
^7{\rm Li}+\nu_{e}$  is an important item
in the context of the so called ``Solar neutrino puzzle'' for
several reasons:\\
i) The result of Gallium experiments would be (partially) consistent
with the hypothesis of standard neutrinos only if  Beryllium neutrino
($\nu_{Be}$) production rate, $L(\nu_{Be})$, 
is suppressed by an order of magnitude with respect 
to the prediction of the Standard Solar Model (SSM), see e.g. 
\cite{where,report}.\\
ii) If one accepts neutrino oscillations as the solution of the
solar neutrino puzzle, the  determination of the neutrino mass matrix 
depends however on the predicted value of $L(\nu_{Be})$.\\
iii) Direct experiments aiming to the determination of the
 $\nu_{Be}$ signal are in preparation and of course
the interpretation of their result will rely 
on  $L(\nu_{Be})$ \cite{borexino,lens}.

The SSM prediction for Beryllium neutrinos \cite{PLB98},
\begin{equation}
\label{eqlbessm} 
L(\nu_{Be})^{SSM} =1.3 \cdot 10^{37} \, (1\pm 9\%) 
{\mbox{ s$^{-1}$  ($1\sigma$ error)} }
\end{equation}
is very robust, much more than that of Boron neutrinos. However
any additional information which does not rely on SSM are
clearly welcome.
In a previous paper \cite{RV99} we have shown that helioseismology,
supplemented with the Superkamiokande result on $^8$B neutrinos,
already yields a lower limit for $^7$Be production:
\begin{equation}
\label{eqlimit}
L(\nu_{Be})\geq 2.8 \cdot 10^{36}\, (1\pm 24\%) 
\,{\mbox{s$^{-1}$}} ~.
\end{equation}
In this paper we make a step forward and we provide a determination
of $L(\nu_{Be})$ directly by means of helioseismology without using 
additonal assumptions.

The basic idea is the following. In the SSM the pp-II termination 
(which is the Beryllium production branch)
accounts for an appreciable fraction of the 
$^4$He produced near the solar center.
As a consequence, if Beryllium  production were suppressed 
-- now and in the past -- less $^4$He would have been produced near
the center. As a consequence,
the molecular weight there decreases and one expects that
the sound speed increases in this region.
In other words, we know that SSM calculations are in 
good agreement with helioseismology and we can expect 
that this agreement is spoiled if $\nu_{Be}$ production 
is substantially altered.

In refs. \cite{Bahc34}  it was shown that models 
where the production of $\nu_{Be}$
is (artificially) forbidden are in conflict
with helioseismology. In this paper we attempt to  a quantitative
determination of the $^7$Be-neutrino  luminosity.
In order to fulfill this program, we have constructed solar models where
$\nu_{Be}$ production is artifically changed  by varying in an arbitrary way
the zero energy  astrophysical S-factor ($S_{34}$)
for the reaction $^3{\rm He}+^4{\rm He}\rightarrow ^7{\rm Be}+ \gamma$. 
As well known (see e.g. \cite{CDF,report,libro}), 
the production rate $L(\nu_{Be})$ 
is directly proportional to $S_{34}$, so that this is an efficient way for 
arbitrary variations of $L(\nu_{Be})$.
We remind that $S_{34}$ is measured with an accuracy of about ten per cent,
($S_{34}^{SSM}=0.54 \pm 0.09$ KeVb \cite{exps34}). 
Its variation well beyond the
experimental uncertainty is just a way of simulating 
several effects which have
been claimed to suppress $\nu_{Be}$ production, e.g. hypothetical 
plasma effects which could alter nuclear reaction rates. These effects, which
correspond to an anomalous screening effect, can be described by introducing
an effective zero energy astrophysical factor $S_{34}\neq S_{34}^{SSM}$.

 
\section{Results and discussion}
\label{secres}

As well known, helioseismology determines quite accurately
several properties of the sun,
see e.g. \cite{eliosnoi,nutel99,Bahcall99}:\\
a) The depth of the convective envelope $R_b$
and the photospheric helium abundance $Y_{ph}$
are determined as:
\begin{equation}
R_b/R_\odot=0.711\pm 0.001
\end{equation}
\begin{equation}
 Y_{ph}=0.249\pm 0.003 \quad,
\end{equation}
where here and in the following the so called ``statistical'' or 1$\sigma$
uncertainties are considered \cite{eliosnoi}.\\
b) The sound speed profile is determined with an accuracy
of about 0.2\% in the intermediate solar region, 
say between 0.1-0.7$R_\odot$.
Recent SSM calculations, see e.g. \cite{PLB98}, 
are in good agreement with these observational data 
within the quoted errors (with the possible exception of the sound speed
just below the convective envelope, which is slightly underestimated
in the calculated models \cite{eliosnoi,PLB98}).

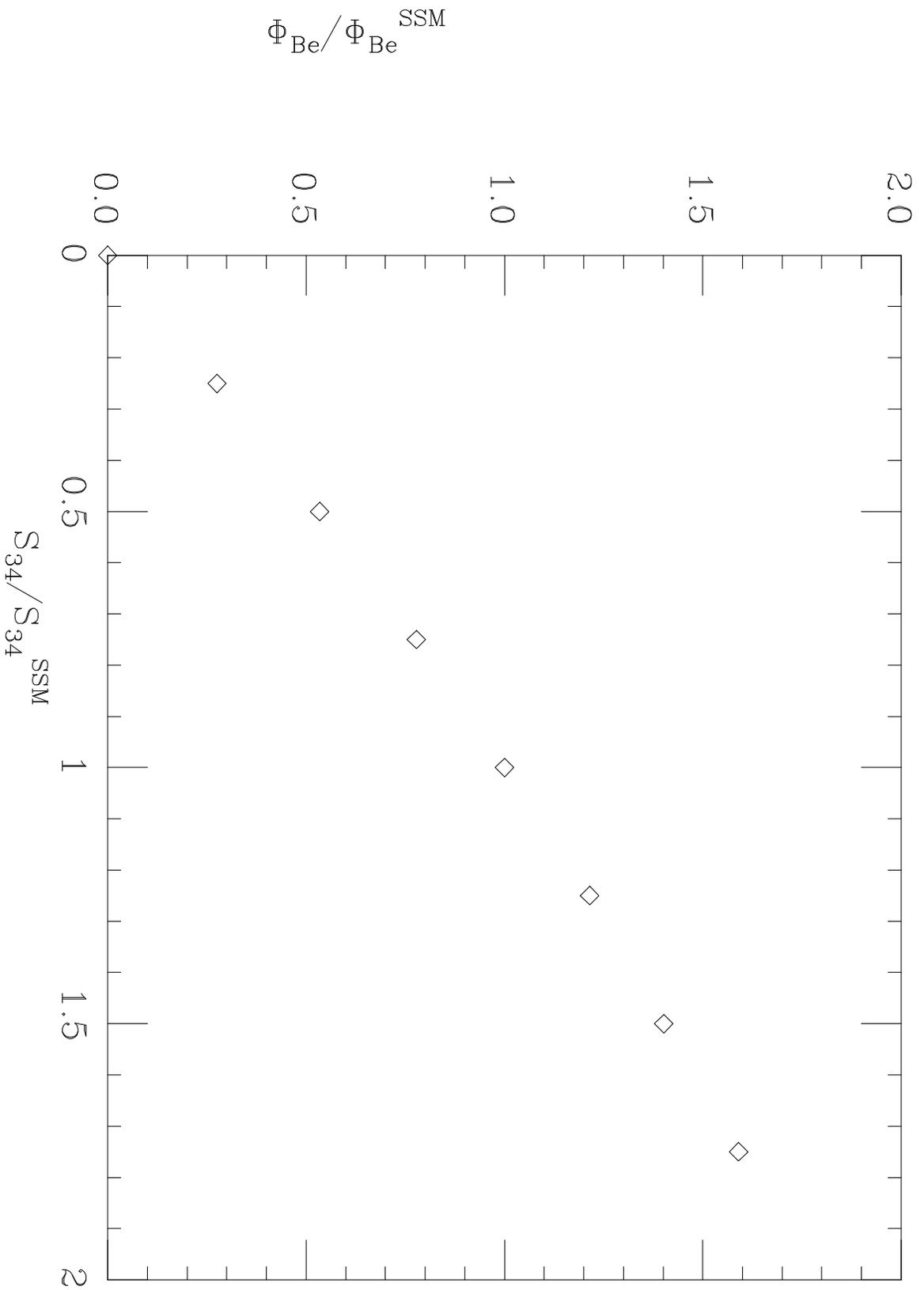
\begin{figure}
\caption{Flux of $^7$Be-neutrinos, normalized to the SSM value,
as a function of $s=S_{34}/S_{34}^{SSM}$.}
\label{figflussi}
\end{figure}

By keeping as a free parameter $s=S_{34}/S_{34}^{SSM}$ we have built a 
series of solar models. This means that, for a fixed value for $s$, 
the stellar evolution code FRANEC \cite{Ciacio}
was run by varying the three free parameters of the model (initial
helium abundance $Y_{in}$, initial metal abundance $Z_{in}$
and mixing length $\alpha$) until it provides a solar structure ,
i.e. it reproduces the observed solar luminosity, radius
and photospheric metal abundance  at the solar age.
As well known by varying $S_{34}$ 
the $\nu_{Be}$ luminosity scales linearly,
see e.g. \cite{CDF,report,libro} 
and Fig. \ref{figflussi}.
In fact,  the production rate of Be nuclei is obviously proportional
to $S_{34}$ and  practically each beryllium nucleus produces 
one $\nu_{Be}$ 
(neglecting the small probability of proton capture compared with
that of electron capture).

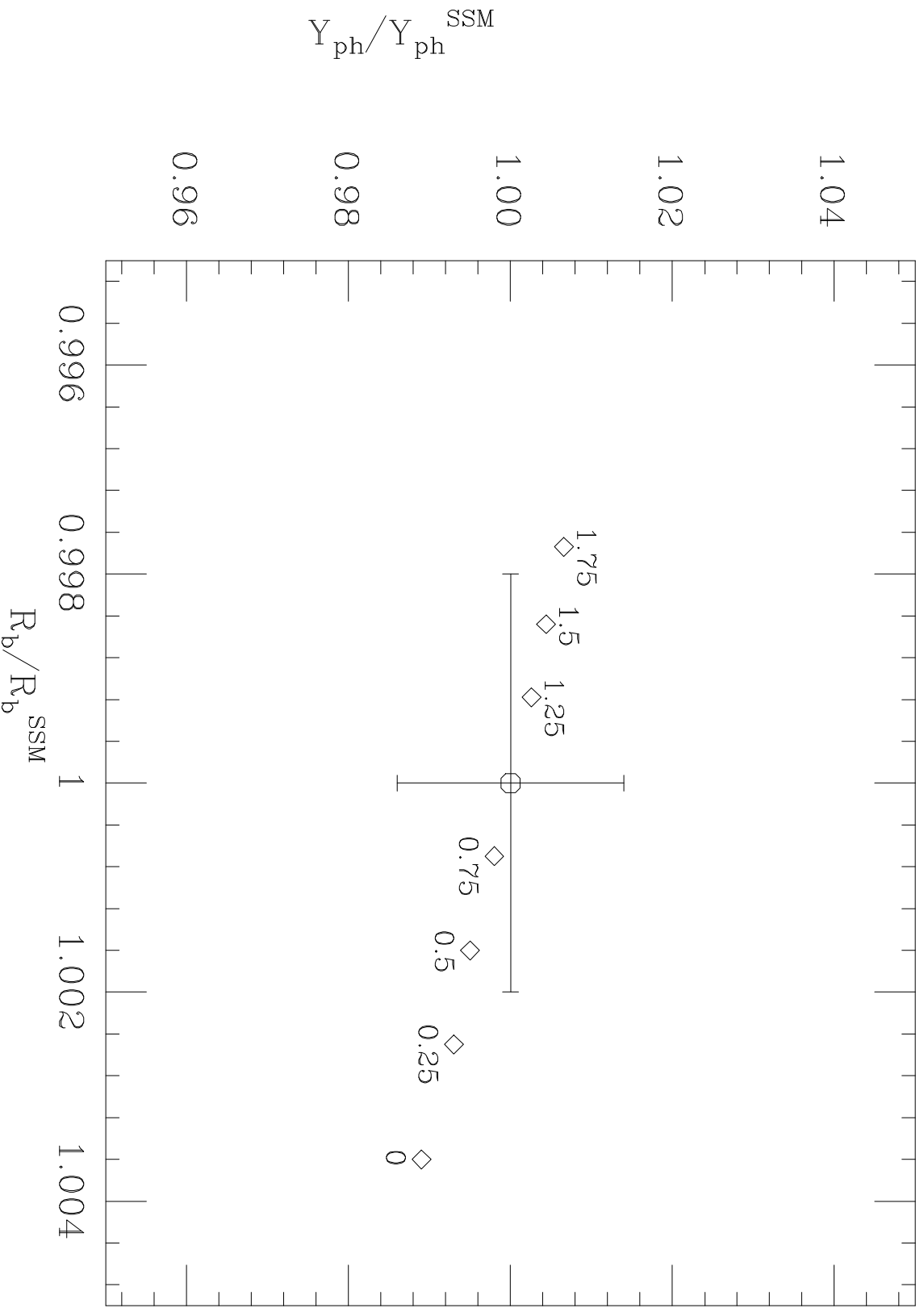
\begin{figure}
\caption{The photospheric helium abundance $Y_{ph}$ and  
the depth of the convective envelope $R_b$ 
in solar models with the indicated
values of $s=S_{34}/S_{34}^{SSM}$.
The error bars  corresponds to the 
$1\sigma$ helioseismic uncertainties, from ref.[11].}
\label{figrb}
\end{figure}

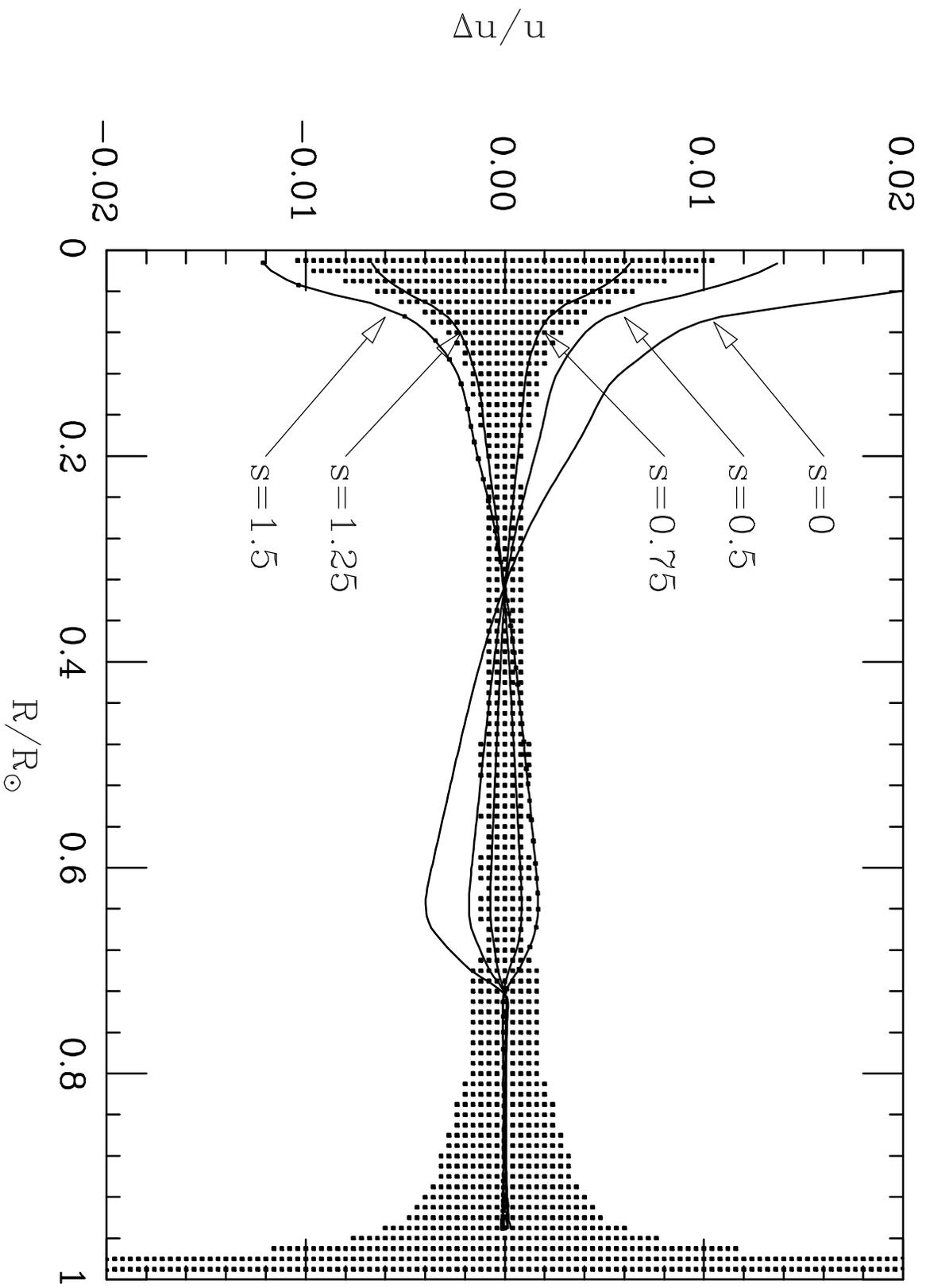
\begin{figure}
\caption{Fractional difference with respect to the SSM
prediction, (model-SSM)/SSM, of the isothermal squared sound
speed, $u=P/\rho$, in solar models with the indicated  
values of $s=S_{34}/S_{34}^{SSM}$. The dotted area corresponds
to the $1\sigma$ helioseismic uncertainty on $u$, from ref. [11].}
\label{figu}
\end{figure}

The resulting values for the quantities which can be tested 
by means of helioseismology are shown in Fig. \ref{figrb}
and Fig. \ref{figu}.

The photospheric helium abundance $Y_{ph}$ is weekly sensitive
to the value of $S_{34}$ whereas the depth of the convective
envelope $R_b$ is altered 
by more than $1\sigma$ if $S_{34}$ is reduced below
one half of the SSM value, i.e.: 
\begin{equation}
L(\nu_{Be})=1.3 \cdot 10^{37} (1\pm 50\%)\quad 
{\mbox{at $1\sigma$}} ~. 
\end{equation}

As previously mentioned, one expects that the sound speed is altered, 
particularly near the solar center. In fact, stringent constraints 
arise from the sound speed profile,
particularly near $R\simeq 0.2 R_\odot$.
The requirement that the sound speed is not changed 
by more than $1\sigma$ yields:
\begin{equation}
L(\nu_{Be})=1.3 \cdot 10^{37} (1\pm 0.25)\quad
 {\mbox{at $1\sigma$}}~. 
\end{equation}

In conclusion, 
helioseismology directly confirms
the production rate of  Beryllium neutrinos as
predicted by SSMs to within  $\pm 25\%$ ($1\sigma$ error).
This constraint is somehow weaker than 
that  estimated from uncertainties of the SSM,
see Eq. \ref{eqlbessm}, however it relies on direct observational 
data.

Finally we remark that the helioseismic determination of
$^7$Be-neutrino production rate also shows that the sun is
producing $^7$Be nuclei and that the $^7$Be abundance is correctly
calculated by SSMs.

\acknowledgments
We are extremely grateful to V. Castellani, 
S. Degl'innocenti and G. Fiorentini for useful suggestions 
and comments.

\end{document}